\documentstyle[12pt, epsfig]{article}
\begin{document}
\def\strut{\rule[-.5cm]{0cm}{1cm}}
\def\dspace{\baselineskip = .30in}

\title{
\begin{flushright}
{\large\bf CERN-TH/96-129}
\end{flushright}
\vspace{1 cm}
{\Large\bf Natural Inflation in SUSY and Gauge-Mediated Curvature\\
of the Flat Directions}}

\author{{\bf Gia Dvali}\thanks{
E-mail:dvali@surya11.cern.ch
}\\ Theory Division, CERN,\\ 
Geneva, Switzerland\\}

\date{ }
\maketitle

\begin{abstract} 
Supersymmetric theories often include the non-compact 
directions in the field space along which the tree level
potential grows only up to a certain limited value (determined by the
mass scale of the theory) and then stays constant for the
arbitrarily large expectation  value of the field parametrizing the
direction. Above the critical value, the tree-level
curvature is large and positive
in the other directions. Such plateaux
are natural candidates for the hybrid inflaton. The non-zero 
$F$-term density along the plateau spontaneously breaks SUSY
and induces the one-loop logarithmic slope 
for the inflaton potential. The coupling of the inflaton
to the Higgs fields in the complex representations
of the gauge group, may result in a radiatively
induced Fayet--Iliopoulos 
$D$-term during inflation, which destabilizes some of the squark and
slepton flat directions. Corresponding soft masses can be
larger than the Hubble parameter and thus, play a crucial role for the
Affleck--Dine baryogenesis.

\end{abstract}
\newpage

\dspace
 
\subsubsection*{Introduction}
The usual problem with inflationary scenarios \cite{linde}
is that they require
introduction of the unmotivated small parameters: the scalar field
with an extremely small self-coupling and the flat potential.
In the non-supersymmetric theories
the scalars with such a flat potential do not emerge
naturally \footnote{Unless the inflaton is a pseudo-Goldstone 
particle\cite{natinf}}.
On the contrary, it is well known that the supersymmetric
theories often admit the flat vacuum directions, which are flat to
all orders in perturbation
theory in the unbroken SUSY limit.
They gain a very small
curvature (typically $\sim$ 100 GeV), induced by supersymmetry breaking
and, therefore, inflation along these directions would be marked with
a very small value of the Hubble constant\footnote{Here and below we
are talking about the vacuum energy driven inflation. The situation
can be very different in the case of the
kinetic inflation, e.g. in string cosmology
\cite{kinetic}} \cite{flat}.
Such an inflation can probably, at best,
be considered on top of the earlier inflation with a sufficiently large
Hubble parameter ($\sim 10^{13}$ GeV or so)
motivated by the large density perturbations consistent with COBE data
\cite{COBE}. However, even if not
responsible for the inflation,
the supersymmetric 
flat directions can still play a very important role in cosmology. 

The aim of the present paper is twofold: first, to
point out some natural candidates for the inflation
that involve a large mass scale
(perhaps motivated from the particle physics), e.g.
such as the GUT scale $M_G$, and secondly to discuss the behaviour of the
flat directions during such an inflation.
We will argue 
that supersymmetric theories often include
certain `plateau' directions in the field space, parametrized by the
scalar fields $X$,
which in the global minimum have large positive curvature
induced by large VEVs of the other field(s) $\phi$. 
Nevertheless, the potential
along such a direction grows up to a certain value and then becomes
frozen, since for large values of $X$ the $\phi$ VEV is switched-off
and the tree-level potential is exactly flat for $ X \rightarrow \infty $.
Since for a large $X$  the potential is non-zero, 
supersymmetry is broken and there is a one-loop induced effective potential
that drives $X$ back to the SUSY minimum. This scenario leads to a natural
realization of the `hybrid inflation' idea
invented by Linde \cite{hybrid}.

 During inflation the flat directions, which have no tree-level couplings
with an inflaton, gain masses $\sim H$ \cite{ddrt}
through the gravity-mediated
supersymmetry breaking \cite{bfs}.
The gauge-charged
flat directions, however, can gain the gauge-mediated soft masses
if the inflaton
interacts with some gauge-non-singlet fields \cite{dv}.
It is well known \cite{gm}
that the gauge interaction also can be a messenger of SUSY breaking,
and in fact more efficient than gravity, provided the messenger scale is
lower than $M_{Planck}$.
This is precisely
what is going to happen in the inflationary scenario discussed above, since
the inflaton has renormalizable couplings with gauge nonsinglet
fields $\phi$ and itself also can be a gauge non-singlet field, e.g.
can carry colour and electric charge. Thus, during inflation the
$\phi$-fields play the role of messengers of supersymmetry
breaking and radiatively induce the universal
(up to charges) two-loop soft masses for all the
gauge non-singlet flat directions, which have no tree-level couplings
with the inflaton. The important thing is that these soft masses
can be greater than $H$, typical
magnitude of the gravitationally induced soft masses.
 Here we observe that when the inflaton couples to the fields in the
complex representations, there is a one-loop-induced 
Fayet--Iliopoulos D-term during inflation. This term can dominate
both the gauge-mediated two-loop and the gravity-mediated soft
terms, and destabilize some of the
squark/slepton flat directions during inflation. Therefore it can
play a crucial role for the  Affleck--Dine mechanism of baryogenesis
\cite{ad}.

\subsubsection*{Simple Examples}

The simplest inflationary scenario with the above properties
was considered in \cite{dss}
The superpotential of the model is\footnote{A similar superpotential was
considered in \cite{cetal}, but the inflaton one-loop
effective potential, which plays
a crucial role, was ignored there.}
\begin{equation} 
W = {1 \over 2} fX\phi^2 - X\mu^2,  
\end{equation}
where $\phi$ is a superfield that breaks the GUT symmetry and 
$\mu$ is a mass scale such that $\mu \sqrt{2 \over f} = M_G$;
$\phi$ can be a component of the Higgs field in the real (adjoint) or
complex representation, e.g. a spinor of $SO(10)$\cite{rj}
or a $6$-plet of $SU(6)$ (see below). 
The above superpotential is the most general one compatible with
the GUT symmetry and an $R$-symmetry under which $X$ carries one
unit of the superpotential charge.
The scalar potential is given by
\begin{equation}
V = \left |{1 \over 2}f\phi^2 - \mu^2 \right |^2 + 
f^2|X|^2|\phi|^2 + D-{\rm terms}
\end{equation}
This theory has a unique supersymmetric
vacuum with $\phi^2 = {2 \over f}\mu^2$
and $X=0$. However, minimization with
respect to $\phi$ for the fixed values of $X$ shows
that for $X > X_c = {\mu \over \sqrt{f}}$, the minimum is at $\phi = 0$,
the potential is flat in the $X$ direction and has a large curvature 
($\sim f|X|$) in
the $\phi$ direction.
This system, under the assumption of the chaotic initial conditions
with $|X| >> X_C$
naturally leads to the inflation.

 Since the curvature in the $\phi$ direction is very large,
we expect that $\phi$ will rapidly settle in its instant minimum
with $\phi = 0$.
In contrast, the curvature in the $X$ direction is zero and
the system will evolve towards the global minimum very slowly.
This state is dominated by large
$|F_x| = \mu^2 $ term, which leads to the inflation.
The one loop-corrections
\begin{equation}
\Delta V = {(-1)^F \over 64\pi^2} TrM^4{\rm ln}{M^2 \over \Lambda^2}
\end{equation}
provide non-zero curvature driving $X$ towards the SUSY vacuum.
The  one-loop corrected effective potential
for the large $X$ behaves as\cite{dss}
\begin{equation}
V = \mu^4 \left ( 1 + {f^2 \over 16\pi^2}
\left [2{\rm ln}{f^2|X|^2 \over \Lambda^2} + 3\right ] \right ).
\end{equation}
The phase transition
with gauge symmetry breaking takes place only after 
the $X$ field drops to its
critical value $X_c$. Below this point, all the VEVs rapidly adjust to
their supersymmetric values.

Since the $F_x$-term, 
which dominates the inflationary Universe,
splits the Fermi--Bose masses of the gauge-non-singlet superfield(s) 
(in the above case $\phi$), there are radiatively induced two-loop soft
masses of all gauge-non-singlet
flat directions (and in particular squarks and
sleptons).
In this case $\phi$ plays the role of the messenger
of the inflaton-induced SUSY breaking for all other gauge-non-singlet
fields. Integrating out the heavy messengers at each point of inflationary
trajectory, we end up with a two-loop ($X$-dependent) soft masses
for the light scalars\cite{dv}
\begin{equation}
m^2_{soft} \sim \left ( {\alpha \over 4\pi} \right )^2
{\mu^4 \over |X|^2}.
\end{equation}
These masses can be larger than the Hubble
constant:
\begin{equation}
H^2 = {\mu^4 \over 3M^2}
\end{equation}
where $M = {M_{Planck} \over \sqrt{8\pi}}$.
Thus, gauge-mediated corrections can be the dominant source
of the scalar soft masses during inflation and cannot be neglected.

The sign of the soft masses
plays a crucial role for the baryogenesis via the 
Affleck-Dine mechanism, since this
mechanism requires large expectation
values of squarks and sleptons along the flat directions.

 Here we wish to show that in more generic (and realistic) cases
there are similar one-loop corrections with either sign.
They appear due to the one-loop-induced Fayet--Iliopoulos $D$-term
during inflation. This generically happens in the theories
in which the inflaton couples to the Higgs fields
in the complex representation that is required for
lowering the rank of the
group in all GUTs other than $SU(5)$. These corrections tend to
destabilize the flat directions during inflation and, thus,
naturally lead to the Affleck--Dine mechanism of the baryogenesis.

 Let us consider a minimal structure that leads to such a picture.
We introduce a pair of Higgs $\phi_+, \phi_-$ fields with opposite
charges under a certain $U(1)$-group. We will think of this $U(1)$
as being a broken Abelian subgroup of some GUT symmetry under which
$\phi_+,\phi_-$ transform in the complex representations. We assume
that in the tree-level global minimum $U(1)$ is broken by the VEV of 
$\phi_+,\phi_-$ fields (along the $D$-flat direction
$\phi_+ = \phi_- = M_G$) triggered by the singlet $S$.
Furthermore, we assume that this VEVs give mass to some charged
states $A_-, A_+$ by mixing them with the neutral one $X$.
The simplest superpotential with above properties can be
chosen as
\begin{equation}
W = fS\phi_+\phi_- + \mu^2S + X(a_+A_+\phi_- + a_-A_-\phi_+).
\end{equation}
On top of this we assume that there are
some U(1)-charged flat directions
that do not have direct couplings
with the above fields in the superpotential.
Again, this superpotential is the most general under $U(1)$ and the
$R$-symmetry under which $S$ carries one unit
of the superpotential charge.
The scalar potential of this system reads
\begin{eqnarray}
V &=& \left | f\phi_+\phi_- - \mu^2 \right |^2 + 
\left | a_+A_+\phi_- + a_-A_-\phi_+ \right |^2 + 
|a_+X\phi_-|^2 + |a_-X\phi_+|^2 \nonumber\\
&+& \left | fS\phi_+ + a_+A_+ X \right |^2 + 
\left | fS\phi_- + a_-A_- X \right |^2 \nonumber\\
&+& {g^2 \over 2}\left ( |\phi_+|^2 -  |\phi_-|^2 +
|A_+|^2 - |A_-|^2 + q_i|Q_i|^2 \right )^2,
\end{eqnarray}
where the $Q_i$ are all other charged fields and $q_i$ are their charges.
This potential has a supersymmetric minimum with
\begin{equation}
\phi_+ = \phi_- = {\mu \over \sqrt f}~~~
S = X = A_+ = A_- = 0.
\end{equation}
In addition, we assume that there are flat directions along which some
combinations of the $Q_i$ fields are undetermined. In this vacuum
one combination ${a_+A_+  + a_-A_- \over \sqrt{a_+^2 + a_-^2}}$
is mixed with $X$ and
gets mass $\sqrt{a_+^2 + a_-^2 \over f }\mu$.
Minimizing the potential for the different fixed values of 
$X$ we find that for $|X| > X_c = \sqrt{ f \over a_-a_+}\mu$
the minimum is at 
$A_+ = A_- = \phi_- = \phi_+ = 0$
and $S$ undetermined. Thus, the potential has a fixed value
$\mu^2$ and a vanishing curvature
in the $X$ direction for any $|X| > X_c$.
In addition the curvature in the
$S$ direction is also zero, since $F_s$ breaks SUSY and the scalar
component is a partner of the Goldstone fermion.
As before, the one-loop corrections provide logarithmic
slope for the $X$ and $S$ and drive them towards the global minimum.
For simplicity we will quote the form of the effective
potential along the
$S = 0$. For $ S = 0 $ and $|X| >> X_c$ the contribution to the
one-loop effective potential only comes from the 
superfields $\phi_-,\phi_+$
whose masses are split by the non-zero $F_s$-term. This superfield
delivers two massive fermions with masses $a_+|X|$ and $a_-|X|$
and two complex scalars with mass-squared
\begin{equation}
m^2_{\pm} = {|X|^2 \over 2} \left [ (a_+^2 + a_-^2) \pm (a_+^2 - a_-^2)
\sqrt{ 1 + {4f^2\mu^4 \over |X|^4(a_+^2 - a_-^2)}} \right ].
\end{equation}
For the large $X$ the one-loop-corrected effective potential
behaves as
\begin{equation}
V = \mu^4 \left ( 1 + {f^2 \over 16\pi^2} 
\left [ 1 + 2 {(a_+^2lna_+^2 - a_-^2lna_-^2) \over (a_+^2 - a_-^2)} + 
2ln{|X|^2 \over \Lambda^2}\right] \right ).
\end{equation}
The crucial difference between this scenario and the one
discussed before is that now, during
inflation, one-loop radiative corrections
induce the Fayet--Iliopoulos $D$-term for the $U(1)$ gauge superfield,
which will be compensated by some of the flat directions in the $Q_i$
sector that carry the $U(1)$ charge of the appropriate sign.
The reason for the appearance of this $D$-term is that during
inflation the messengers $\phi_+$ and $\phi_-$
have different masses, since
$a_+^2 \neq a_-^2$, and their contributions do not cancel out in the
one-loop diagram (see Fig.1).
\begin{figure}
\epsfig{file = 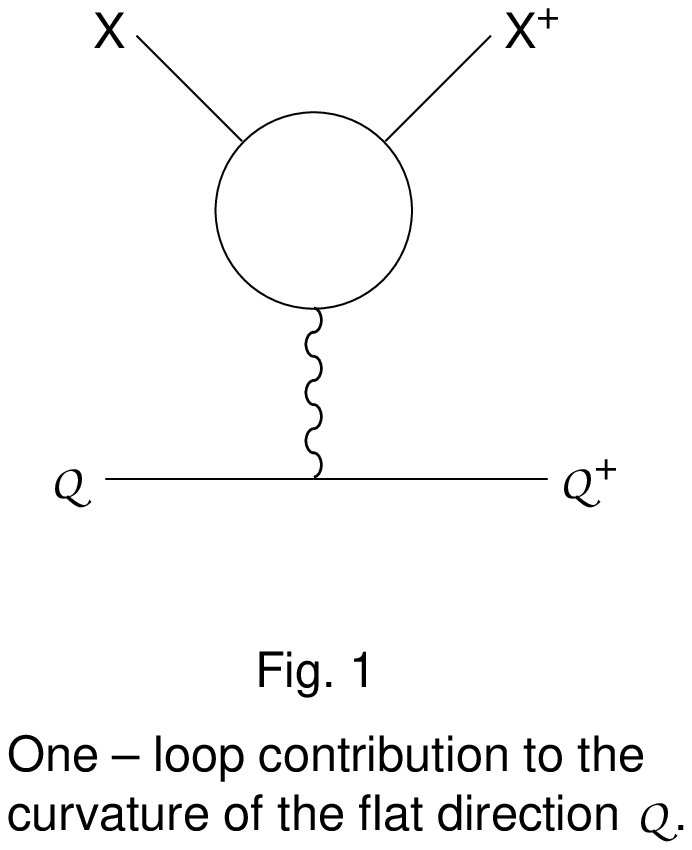, width = 9cm}
\end{figure}
In the leading order, the
corresponding soft masses for the light scalars
are proportional to
\begin{equation}
q_i \left ({\alpha \over 4\pi} \right )
{f^2\mu^4 \over |X|^2} \left [ {a_+^2 + a_-^2 \over (a_+^2 - a_-^2)^2}
ln{a_+^2 \over a_-^2} + {2 \over a_+^2 - a_-^2} \right ] \sim
H^2 {M^2 \over |X|^2} \left ({\alpha \over 4\pi} \right ).
\end{equation}
Now, the inflation ends when the slow roll conditions break down,
that is when either $\left |{M V' \over V }\right | \sim 1$
or $|V''| \sim H^2$ \cite{linde},
where the prime denotes the derivative in the inflaton
direction. For the inflaton potential given by (11), this happens
when $X$ drops to the value $|X| \sim \sqrt{C}M$, where $C \sim 
{f^2 \over 8\pi^2}$ is a loop factor. 
Thus, at the end of the inflation the gauge-mediated soft masses are
$\sim H^2C^{-1}\left ({\alpha \over 4\pi} \right )$. In the interval
$X_c < |X| < \sqrt{C}M$, between the end of inflation and the
gauge-symmetry-breaking phase transition, the ratio of the
gauge-mediated to gravity-mediated soft masses is
\begin{equation}
{m_{gauge}^2
\over m_{gravity}^2} \sim {M^2 \over |X|^2} \left ({\alpha \over
4\pi } \right).
\end{equation}
We see that (at least on the later stages of the inflation)
the gauge-mediated corrections are dominant and can thus
play the crucial role for the Affleck--Dine
scenario of the baryogenesis.

\subsubsection*{Realistic Example}

 The inflationary scenario outlined above can naturally emerge in many
realistic GUTs in which the Higgs fields in the complex representation
are lowering the rank of the group.
Here we consider one example motivated from particle physics.
This is an $SU(6)$ GUT in which the doublet--triplet splitting problem
is solved by the pseudo-Goldstone mechanism \cite{su(6)}. 
The minimal Higgs structure, group theoretically required for the breaking
of $SU(6)$ down to $ G_W = SU(3)_c\otimes SU(2)_L\times U(1)_Y$ 
is one adjoint $35$-plet  $\Sigma_i^k$ and a pair of fundamental and
anti-fundamental representations ($6,\bar 6$-plets) $\phi_i$ and
$\bar{\phi}^i$ ($i,k = 1,2,.., 6$ are $SU(6)$ indexes).
This minimal structure suffices to deliver a pair of light
electroweak Higgs doublets and naturally solve the doublet--triplet
splitting problem. The crucial assumption is that $\Sigma$ and
$\phi, \bar{\phi}$ have no direct cross couplings,
in the superpotential,
so that the Higgs superpotential has the form:
\begin{equation}
 W_{Higgs} = W_{\Sigma}(\Sigma) + W_{\phi}(\phi, \bar{\phi}).
\end{equation}
The absence of the possible cross term $\bar{\phi}\Sigma\phi$ can
be guaranteed by the exact discrete or continuous symmetries of the
theory, which we will not specify here; just note that the
one possible candidate is an $R$-symmetry 
(for the concrete realizations see
\cite{su(6)},\cite{su6}).
For the breaking of the gauge group to $G_W$ the $\Sigma$
and $\Phi$-fields should pick up the VEVs in the 
$SU(4)\otimes SU(2)\otimes U(1)_q$-invariant and 
$SU(5)$-invariant directions respectively:
\begin{equation}
\Sigma = {\rm diag}(1,1,1,1,-2,-2)\sigma ~~~
\phi= (\phi_+,0,0,0,0,0)~~~\bar{\phi}= (\phi_-,0,0,0,0,0).
\end{equation}
Due to the absence of the cross couplings, the Higgs superpotential
has an $SU(6)_{\Sigma}\otimes SU(6)_{\phi}$ global symmetry, which
gets broken to 
\begin{equation}
G_{global} = \left [ SU(4)\otimes SU(2)\otimes U(1)_q
 \right ] _{\Sigma} \otimes \left [ SU(5) \right ] _{\phi}.
\end{equation}
Simple counting of the Nambu--Goldstone modes shows that there is
a pair of massless (in the SUSY limit) electroweak doublet states,
which are not eaten up by the gauge fields and are physical Higgs
particles.

 In this model quarks and leptons are placed in the 
$15^{\alpha} + \bar{6}^{\alpha} +\bar{6}^{'\alpha}$
representations ($\alpha = 1,2,3$ is a family index),
which under the  $SU(5)$ group decompose as 
\begin{equation}
10^{\alpha} + 5^{\alpha} + \bar{5}^{\alpha} +
\bar{5}^{'\alpha} + 1^{'\alpha} + 1^{\alpha}
\end{equation}
and we see that there are on top of the usual chiral
matter (quarks and leptons in
three $10 + \bar 5$ copies),
extra vector-like states $5 + \bar 5$ and
two singlets per family (one combination of singlets can play the role
of the right-handed neutrino). These extra states gain GUT scale
masses from the VEV of $\phi, \bar{\phi}$. The fermion masses in this
GUT were studied in detail in \cite{fermions} and the interested reader
is referred to this paper. Here we will concentrate on the part
of the superpotential that is responsible for the
breaking $SU(6) \rightarrow SU(5)$ and 
for the generation of the heavy matter masses.
The simplest possible form for this sector can be chosen as
\begin{equation}
W = fS\phi\bar{\phi} - S\mu^2 + a_{\alpha}\bar{\phi}15^{\alpha}
\bar 6^{\alpha} + b_{\alpha}\phi\bar 6^{'\alpha}X_{\alpha}. 
\end{equation}
where
$X_{\alpha}$ are three additional singlets. Here we have introduced
a matter parity under which only matter multiplets change sign and
another $Z_2$ symmetry under which only $\bar 6'$ and $X$
superfields change sign.
Without loss of generality, we
have diagonalized Yukawa coupling matrices by the field redefinition.
The potential of this system reads
\begin{eqnarray}
V &=& \left |f\phi\bar{\phi} - \mu^2 \right |^2 + 
a_{\alpha}^2 \left | \bar{\phi}\bar 6^{\alpha} - 
\bar 6^{\alpha}\bar{\phi} \right |^2 \nonumber\\
&+& a_{\alpha}^2\left | \bar{\phi}15^{\alpha} \right |^2 +
b_{\alpha}^2 \left ( |\phi|^2|X_{\alpha}|^2
+ |\phi\bar 6^{'\alpha}|^2\right )\nonumber\\
&+& \left|fS\bar{\phi} + b_{\alpha}\bar 6^{'\alpha}X_{\alpha}\right|^2 +
\left|fS\phi + a_{\alpha}15^{\alpha}\bar 6^{\alpha} \right|^2 +
D-terms
\end{eqnarray}
($SU(6)$ indices are suppressed).
Let us investigate the behaviour of the above potential along the
non-compact $D$-flat direction 
\begin{equation}
15^1_{ik} = X(\delta_i^1\delta_k^2 - \delta_i^2\delta_k^1)~~~
\bar 6^{'1i} = X\delta_i^2~~~\bar 6^{2i} = X\delta_i^1~~~
X_2 = X
\end{equation}
where $X$ is a parameter. It is not difficult to notice that
this direction is not $F$-flat and
has a large $\sim \mu^2$ positive curvature at the origin $X = 0$.
However, for 
\begin{equation}
|X|^2 > X_c = max \left ( {f\mu^2 \over a_1b_2} ,~
{f\mu^2 \over \sqrt 2a_2b_2}
\right )
\end{equation}
the curvature vanishes for arbitrarily large values of $X$, and
the tree-level potential stays constant $V_o = \mu^4$.
This is because the instant minimum in
all other fields (coupled to the given direction) is at their zero value
as far as $|X| > X_c$. The vanishing of the $\phi,\bar{\phi}$ fields,
which gain large $\sim |X|^2$ positive masses, leads to the non-zero
cosmological constant that drives inflation. During this period
supersymmetry is spontaneously broken by the expectation value
$F_s = \mu^2$, which splits masses of the $\phi, \bar{\phi}$
superfields and induces the one-loop effective potential similar to
the one of (11).
We see that the inflationary scenario discussed in the previous
section emerges naturally in this context. Note that $D$-flatness
conditions do not require $X_2 = |X|$, so that we can treat 
$X_2$ as a second parameter of the plateau. The critical values
above which the curvature vanishes will then be defined by the
condition 
\begin{equation}
|XX_2| >  max \left ( {f\mu^2 \over a_1b_2} ,~
{f\mu^2 \over \sqrt 2a_2b_2} \right )
\end{equation}
and the inflationary trajectory will be
parametrized by both $X$ and $X_2$.
The $D$-flat direction we have chosen is in
no way unique and
one may equally well
choose in (20) other combinations of
$15^{\alpha}, \bar 6^{\beta}, \bar 6^{'\gamma}$ and $X_{\nu}$
with $\alpha \neq \beta$ and $\gamma \neq \nu$. The simple rule is that,
from each operator, only one VEV can participate in the plateau direction,
since in the opposite case some of the $F$-terms can grow unbounded
along this direction.

   Now what about the gauge-mediated soft masses during inflation?
To answer this question, first note that an interesting difference
between this inflation and the simple toy model discussed in the
previous section is that the inflaton parametrizes the gauge-nonsinglet
direction along which the colour and the electric charge
are broken. {\it Per se}, this
offers an interesting scenario of baryogenesis
through the out-of-equilibrium decay of the inflaton.
To see which is the unbroken gauge group
during inflation we have to know the relative orientation of the
$\Sigma$ VEV in the basis in which other VEVs are given by (20).
Precise orientation will be decided by the radiative corrections,
which induce an effective potential for the pseudo-Goldstone modes.
The detailed analysis of this dynamics will not be attempted
here and, for definiteness, we assume that the form of sigma is the one
given by (15). In such a case the unbroken gauge symmetry in the
inflationary epoch would be
\begin{equation}
G_{inf} = SU(2)_C\otimes SU(2)_L\otimes U(1)_{YC} 
\end{equation}
where $SU(2)_C$ is a colour subgroup and the generator of
$U(1)_{YC}$, ${\rm diag}(0,0,1,1,-1,-1)$, is a 
combination of hypercharge and colour.
It is expectable that some of the squark and slepton fields
which are not coupled, at tree level, with the inflaton direction,
will be destabilized by one-loop-induced soft masses proportional
to their charges under the broken generators.
Note that, in this particular
example, there is no contribution
to one-loop soft masses from the
$U(1)_{YC}$ gauge fields. This is because in
the above case $\phi$ and 
$\bar{\phi}$, which play the role of the 
SUSY-breaking messengers during inflation,
are decoupled at tree level from the $\Sigma$, and the 
masses of their $U(1)_{YC}$-charged components are degenerate.
In contrast, the masses of the components
that carry charge under broken
generators (e.g. ${\rm diag}(-5,1,1,1,1,1)$)
are not degenerate. This can be
seen from the tree-level values of these mass terms
along the inflationary trajectory:
\begin{eqnarray}
&&|\bar{\phi_1}|^2a_1^2|X|^2 + |\bar{\phi_2}|^2(a_1^2 + 2a_2^2)|X|^2 +
|\bar{\phi_k}|^22a_2^2|X|^2 + |\phi_1|^2b_2^2|X_2|^2 + \nonumber\\
&&|\phi_2|^2(b_1^2|X|^2 + b_2^2|X_2|^2) + |\phi_k|^2b_2^2|X_2|^2 +
f\phi\bar{\phi}\mu^2 + {\rm h.c.}
\end{eqnarray}
where $k > 2$. The resulting one-loop contribution to the soft masses
in general is
$\sim \left( {\alpha \over 4\pi} \right ) {\mu^4 \over |X|^2}$.

\subsubsection*{Conclusions}

The aim of the present letter was to point out that the supersymmetric
theories that include a large mass scale may
contain a potential source for the natural inflation --
non-compact plateau directions in the field space along
which the tree-level potential has a vanishing curvature and large
constant value. Such directions may be parametrized by
the gauge-singlet or gauge-non-singlet fields. During inflation
at least some of the chiral
superfields that have tree-level couplings with the
inflaton suffer from the Fermi--Bose mass splitting and effectively
play the role of messengers for the
gauge-mediated supersymmetry breaking.
In the case of the complex representations, this breaking
can generate
the one-loop Fayet--Iliopoulos $D$-term during inflation, which can
destabilize some of the squark/slepton VEVs. The resulting 
gauge-mediated soft masses
are typically larger 
than the gravity-mediated ones ($\sim H$); they can thus play
a crucial role for the Affleck--Dine scenario of baryogenesis.

\subsubsection*{Acknowledgements}

I learned that recently A.Linde and A.Riotto have
discussed in a different
context the relevance of the SUSY breaking at
the end of inflationary epoch \cite{priv}.

\end{document}